\documentclass[12pt]{iopart}

\usepackage{graphicx}
\usepackage{epsfig}
\usepackage{setspace}
\usepackage{wasysym} 
\usepackage{url}
\usepackage{cite}

\begin{document}

\title{On the origin and evolution of icicle ripples}


\author{Antony Szu-Han Chen and Stephen W. Morris}

\address{Department of Physics, University of Toronto, \\
60 St. George St., Toronto, ON Canada M5S 1A7}
\ead{aschen@physics.utoronto.ca, smorris@physics.utoronto.ca}
\begin{abstract}
Natural icicles often exhibit ripples about their circumference which are due to a morphological instability.  We present an experimental study that explores the origin of the instability, using laboratory-grown icicles.  Contrary to theoretical expectations, icicles grown from pure water do not exhibit growing ripples.  The addition of a non-ionic surfactant, which reduces the surface tension, does not produce ripples.  Instead, ripples emerge on icicles grown from water with dissolved ionic impurities.  We find that even very small levels of impurity are sufficient to trigger ripples, and that the growth speed of the ripples increases very weakly with ionic concentration.   
\end{abstract}




\section{Introduction}

\begin{figure}
\begin{center}
\includegraphics[width=15.5cm]{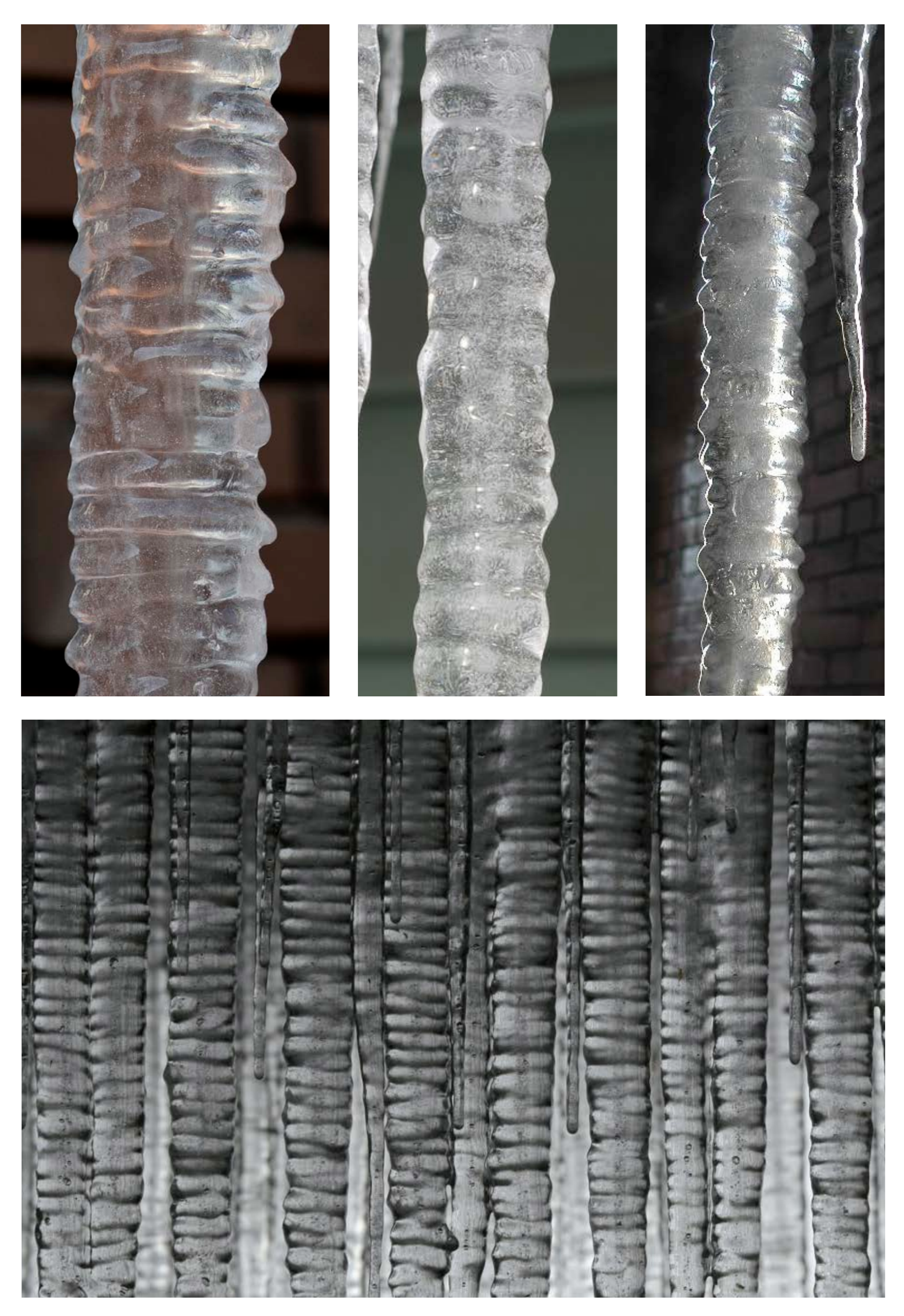}
\caption{Natural icicles that exhibit ripples.  In all cases, the ripple wavelength is close to 1~cm. [ Upper left image, courtesy of Ross McPhail.  Bottom image, courtesy of Jeremy Hiebert ].}
%
\label{natural}
\end{center}
\end{figure}

Consider the icicle. Like many processes far from equilibrium, the growth of icicles can exhibit nonlinear pattern-formation~\cite{cross_greenside}.  Icicles often exhibit a regular pattern of ripples, sometimes called ``ribs", around their circumference~\cite{natural_icicles,maeno_japanese2,sapporo_icicles,chung_salty,maeno,matsuda_thesis,ogawa,ueno1,ueno2,ueno4,ueno5,ueno_ripple_expt,ueno_farz_2011,Chen_PRE}. Some natural examples are shown in Fig.~\ref{natural}.   These ripples are superposed upon a nearly self-similar overall shape~\cite{Chen_PRE,Short_Phys_Fluids} and, like the shape, the ripple wavelength is remarkably universal, independent of the growing conditions.   Ripples have been observed to move slowly 
upward during growth~\cite{maeno,ueno_ripple_expt,Chen_PRE}, and occur on both natural~\cite{natural_icicles,maeno_japanese2,sapporo_icicles} and laboratory-grown~\cite{chung_salty,matsuda_thesis,ueno_ripple_expt,Chen_PRE} icicles.  The familiar elongated form of a ripply icicle is the result of a subtle interplay between the solidification dynamics of ice~\cite{worster_book,davis_book} and the gravity-driven flow of the thin supercooled liquid water film flowing over its evolving surface~\cite{Short_Phys_Fluids,makkonen1}.  The latent heat released by freezing, which controls ice growth, is advected by the water film and ultimately carried away by the surrounding sub-zero air, which is also flowing~\cite{Short_Phys_Fluids,Neufeld_JFM}.  Ripples are presumed to be due to a morphological instability of ice growth in the presence of a thin flowing water film.  
Icicle ripples have a close resemblance to the ripples, known as {\it crenulations}~\cite{stalactite_ripples}, that appear on stalactites.   These also involve growth under a thin flowing water film and travel upward during growth. Like icicles, stalactites have been predicted to have self-similar universal  shapes~\cite{Short_PRL_stalactites}.  Whether there exists some universal mechanism linking these two natural pattern forming systems remains an interesting open question.

Icicles are perhaps the simplest natural manifestation of the general phenomenon of ``wet" ice growth, in which freezing proceeds into supercooled water delivered by some combination of gravity, wind-driven flow, and spray~\cite{wet_ice_growth}.  Mitigating the hazards of ice accretion on aircraft, ships, and other structures is an important branch of engineering~\cite{engineering}.  Even otherwise innocent icicles amount to unwanted reverse lightning rods if they form on high-voltage equipment~\cite{hv_icicles}.  Ripple growth involves important aspects of the morphological stability of wet ice growth that are also relevant to  practical ice accretion problems.

In this paper, we present a comprehensive experimental study of ripple pattern formation on laboratory icicles that sheds new light on their growth dynamics and on the origin of the rippling instability.  Our study reveals the essential but hitherto 
unappreciated role of dissolved ionic impurities in icicle ripple formation.  Pure water icicles, while not perfectly smooth, do not exhibit growing ripples.  The growth speed of ripples on icicles made from very slightly saline water shows a very weak, almost logarithmic dependence on the salt concentration.  We also examined the effects of surfactant and dissolved gases on the formation of ripples; neither caused ripples to grow without dissolved salt also being present. Finally, we discuss how these results fundamentally disagree with theories of the rippling instability proposed so far~\cite{ogawa,ueno1,ueno2,ueno4,ueno5,ueno_ripple_expt,ueno_farz_2011}. We conclude that icicle ripples, despite their deceptively simple form, continue to elude theoretical explanation.


\section{Experiment}

To make laboratory icicles, one must control the ambient temperature, the flow rate of the feed water and its composition, as well as the state of motion of the air surrounding the icicle.  The humidity of the air is also an important factor that must be monitored.  The time evolution of the icicle morphology can then be acquired from digital images using edge detection.  A typical icicle requires many hours to grow.

In addition to ambient temperature and water supply rate~\cite{maeno}, it has previously been shown that water purity and air motion are important to the overall icicle shape~\cite{Chen_PRE}.  Icicles grown from distilled water are significantly smoother and closer to the self-similar shape predicted by  theory~\cite{Short_Phys_Fluids} than those grown from untreated tap water.  Tap water icicles show prominent ripples and other non-ideal morphologies~\cite{Chen_PRE}. Icicles grown in still air tend to develop multiple tips~\cite{Chen_PRE}.  For this reason, in all the experiments we report here, the air was stirred by fans, as discussed below.     

All previous ripple measurements were made on either natural icicles~\cite{natural_icicles,maeno_japanese2,matsuda_thesis} or laboratory icicles grown from tap water~\cite{matsuda_thesis,ueno_ripple_expt,ueno_private_comm}.  In their growth experiments on distilled water icicles, Maeno~\emph{et al.}~\cite{maeno} commented that it was ``difficult'' to produce ripply ones.  In experiments on marine icicles, Chung~\emph{et al.}~\cite{chung_salty} qualitatively observed that ``pronounced ribs" were a characteristic that distinguished saline water icicles from pure water ones.  All these clues point to the importance of the water purity in ripple formation.  Nevertheless, water purity  was not considered in any of the existing theories of the rippling instability~\cite{ogawa,ueno1,ueno2,ueno4,ueno5,ueno_ripple_expt,ueno_farz_2011}.  No previous experiment has quantitatively examined the relationship between water purity and the growth and evolution of icicle ripples.

We compared icicles grown under various conditions of temperature, water supply rate, and air motion, with special emphasis on varying  the composition of the feed water.  We added small quantities of sodium chloride to distilled water to achieve levels of ionic impurities similar to tap water; we also investigated both the role of dissolved gases and the effect of surfactant which reduces the water-air surface tension.  We found that the presence of small amounts of ionic impurities is crucial to the formation of ripples.

\subsection{Apparatus}

The apparatus was previously described in Ref.~\cite{Chen_PRE}. Icicles were grown below a sharpened wooden support suspended inside an insulated, refrigerated box.  The support was rotated, at a speed of four minutes per revolution, to encourage axisymmetry and to allow all sides of the icicle to be imaged.  The air was gently stirred by internal fans in the corners of the box.   The air motion produced by the slow rotation of the support was negligible compared to that produced by the fans.   
The wall temperature was controlled by a commercial bath, which circulated antifreeze through pipes embedded in the walls of the box.  
The feed water supply was delivered by a peristaltic pump to a temperature-controlled nozzle that was slightly off-axis from the rotating support, in order to distribute the water evenly.  The growing icicle was imaged by a digital camera via a slot in the side of the box. 
The rotational position of the support was indexed so that eight reproducible views of the icicle could be imaged on each rotation.

\subsection{Water samples}

Table~\ref{waters} shows the measured physical parameters of the various water samples used. 
The distilled water was supplied in bulk by Canadian Springs~\texttrademark.  Note that, while it is 
much purer than tap water, it is not near the limits of what purity can be achieved.  Nevertheless, this level of purity was 
sufficient to suppress the rippling instability.  The added NaCl was ACS reagent grade, supplied by Sigma-Aldrich.  The 
added non-ionic surfactant was Triton X-100, 
which is t-Oct-C$_{6}$H$_{4}$-(OCH$_{2}$CH$_{2}$)$_{x}$OH, $x$ = 9-10. The Triton X-100 was also supplied by 
Sigma-Aldrich.


The conductivity and composition of the samples were analyzed by the ANALEST lab, Department of Chemistry, University 
of Toronto using standard analytical techniques. 
The surface tensions of distilled water, tap water, and surfactant solutions were 
measured using the capillary tube method at room temperature.  The surface tensions of salt solutions were calculated 
using Ref.~\cite{MIT_seawater}.  

Water sampled from melting natural icicles and source water from nearby roof runoff was collected in Toronto during March, 2013.  This water typically produced ripply icicles whose visual appearance was similar to those made with tap water.  The composition of Toronto tap water is shown in Table~\ref{waters} for comparison, although we do not report data for any icicles made with tap water here~\cite{Chen_PRE}.  Melted natural icicles and their  source water are generally intermediate in purity between distilled water and Toronto tap water.

\begin{table} \renewcommand{\tabcolsep}{0.3mm} 
\begin{center}
\begin{tabular}{||l|l|l|l||l||}
\hline
{\bf Solution }			&		{\bf Main impurities}		&   {\bf Concentration}  	&	{\bf Conductivity}   	&	
{\bf Surface tension}	\\
\hline
Distilled water (DW)		&		Ca$^{+2}$ 			&  0.037 mg/L 	&	$2$~$\mu$S/cm	&	0.072 N/m	
		\\
					&		 K$^+$ 				&  0.026 mg/L 	&							&	\\
					&		Na$^+$ 				&  0.014 mg/L 	&							&	\\
					& 		Ba$^{+2}$  			&  0.003 mg/L	&							&   	\\
\hline
DW + Triton X-100		&		Triton X-100			&  200 mg/L	&	$1.7$~$\mu$S/cm			&	
0.039 N/m			\\
\hline
DW + NaCl (typical)		&		NaCl 		&	 80.0 mg/L		&	$211$~$\mu$S/cm 	&	0.072 N/m\\
\hline
Natural icicle source		&		Ca$^{+2}$		&	2.42 mg/L		&	$30$~$\mu$S/cm	&			 \\
water (typical)				&		Na$^+$		&	1.83~mg/L	&					&	\\
					&		Cl$^-$		&	3.56~mg/L	&					&	\\
\hline
Toronto tap water		&		Ca$^{+2}$ 	&  37.5 mg/L	&	$419$~$\mu$S/cm		&		0.071 N/m	\\
					&		Na$^+$		& 10.7 mg/L	&						&				\\
					&		Mg$^{+2}$	& 8.82 mg/L	&						&	\\
					&		 K$^+$ 		& 1.66 mg/L	&						&	\\
					&		Si$^{+2}$ 		& 1.22 mg/L	&						&	\\
					&		Cu$^{+2}$ 	& 0.46 mg/L 	&						&	\\
					&		SO$_4^{-2}$ 	&  31.60 mg/L	& 						&	\\
					&		Cl$^-$ 		&  25.11 mg/L 	&						&	\\
					&		NO$_3^-$ 	&  0.96 mg/L 	& 						&	\\
					&		F$^-$ 		& 0.41 mg/L	&						&	\\			
					&		 Zn$^{+2}$, V$^{+2}$, Ni$^{+2}$,  	&  $<$ 0.2 mg/L  &	&	\\
					&		Ba$^{+2}$, Fe$^{+2}$, Mn$^{+2}$		&			&			&	\\					
\hline
\end{tabular}
\caption{Measured compositions, conductivities, and surface tensions of the water samples.  Water that produced ripply natural icicles was collected from Toronto roofs in March, 2013.  Toronto tap  water, which produces icicles exhibiting prominent ripples, is shown for comparison. }
\label{waters}
\end{center}
\end{table}

\subsection{Temperature, humidity, and air flow measurements}

The temperature of the water inlet nozzle was measured with a thermocouple and feedback controlled by means of a computer-controlled heater.  It was typically 
maintained at 3.0$\pm$0.3 $^{\circ}$C to prevent freezing of the inlet pipe.  The temperature of the feed water had no important effect on the icicle growth. The rotating support upon which the icicle grew was suspended by stiff wires and was not actively cooled.  All of the heat from 
the growing icicle was transported by advection and diffusion through the surrounding air, by radiation, or by evaporation of water~\cite{Neufeld_JFM}. 
  The latter mechanism depends on the relative humidity, which was continuously monitored during the experiment.  For all 
the data reported here, the relative humidity was 85$\pm$5~\%.

The air inside the icicle-growing box was stirred by eight small computer fans: four at the top corners pointed straight down, and four at the bottom corners pointed straight up.  Air motion affects the overall growth speed of icicles~\cite{maeno} and is also necessary to suppress the tendency of icicles to grow multiple tips~\cite{Chen_PRE}.  Using a digital anemometer, we determined the total air flux from each fan, which could be varied by changing its applied voltage.  The range of flux used was $(0.65 - 1.23)~{\rm m}^3/{\rm min}$. Dividing the volume of the box by eight times this flux gives a characteristic time in the range of $(1 - 2)$ sec --- this is an estimate of the time it takes for the fans to circulate all of the air enclosed in the box once.  Thus, the air in the box may be regarded as well-stirred, save for a thin viscous and thermal boundary layer near the icicle and near the walls of the box. Thermocouple measurements of the air in the well-stirred region show a mean temperature a few degrees warmer than the walls of the box, which are controlled to within $ \pm$0.3 $^{\circ}$C by the circulating bath.

\subsection{Image analysis}

\begin{figure}
\begin{center}
\includegraphics[width=16cm]{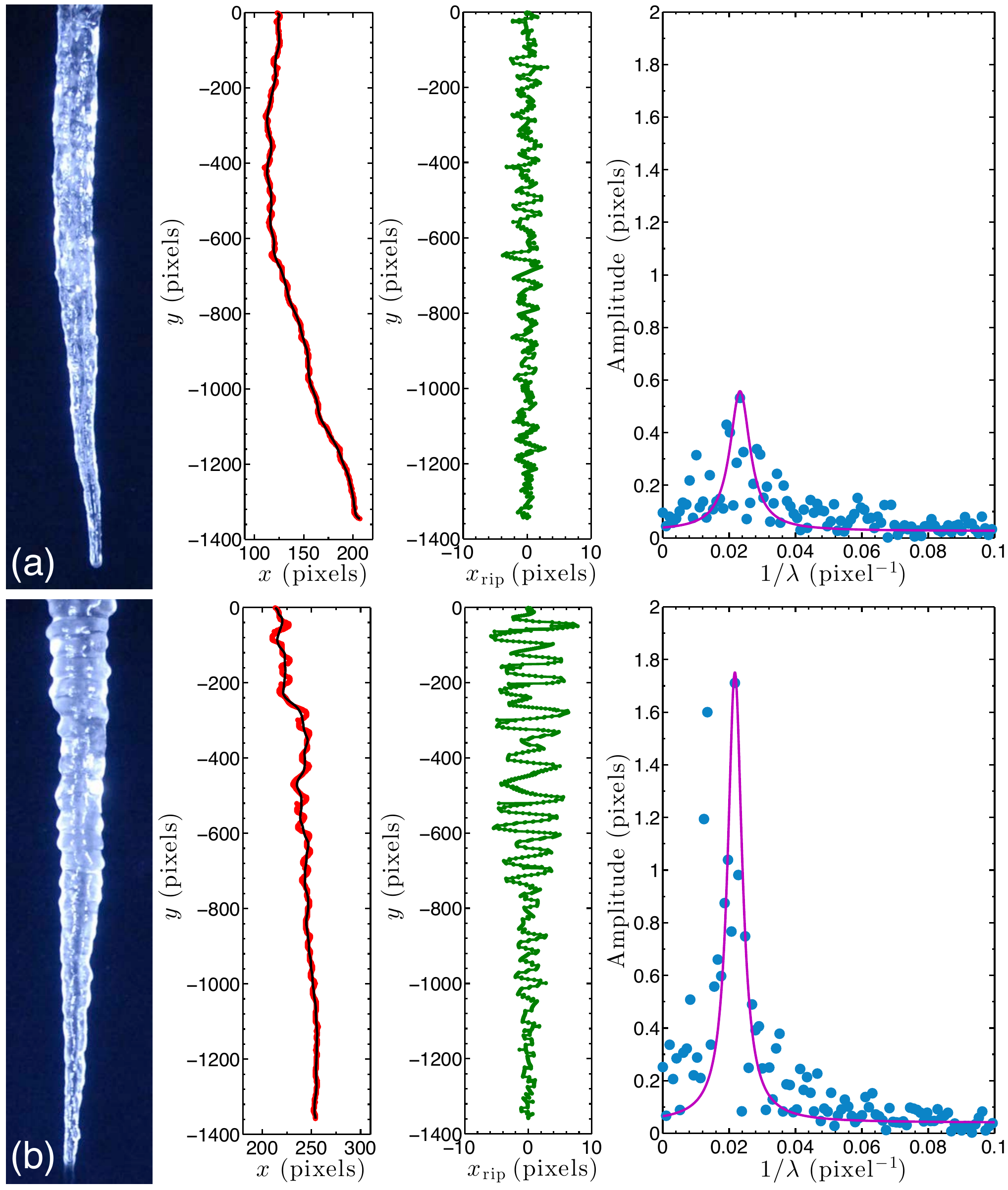}
\caption{ \label{fig_ripalg_NJP}
Ripple extraction and analysis of icicle data: (a) is a distilled water icicle; (b) is an icicle made with distilled water plus $8.0\times10^{-3}$ wt~\% NaCl.  From left to right: the icicle image; the detected left edge (red) and its background shape (black); the ripple position for the left edge (green); the Fourier spectrum of the ripple position for the top 10~cm of the left and right edges, spliced together peak-to-peak (cyan).  The magenta curve in the rightmost plot is a Lorentzian fit through the maximum point of the spectrum. 
}
\end{center}
\end{figure}

To investigate the time evolution of icicle ripples, we analyzed time series of high-resolution digital images taken during the 
experiment.  MATLAB's Sobel algorithm was used to detect the left and right edges of the icicle in each image.  The positions of the two detected edges $x(y)$ were filtered to obtain the ripple positions $x_{\rm rip}(y)$.  Here, $x$ is the horizontal coordinate and $y$ is the vertical coordinate in the image, with $y$  measured downward from the root of the icicle.  The filtering consisted of subtracting a background shape, which was obtained by smoothing the detected edge using a moving average de-trending filter with a vertical span of 1~cm.  The filter de-emphasized topographic features that had longer length scales than typical ripples, which have been shown previously~\cite{natural_icicles,maeno_japanese2,matsuda_thesis,ueno_ripple_expt} to have wavelengths near 1 cm. This choice of de-trending filter is also consistent with the wavelength of all the prominent ripple patterns observed on saline icicles in the present study. Fig.~\ref{fig_ripalg_NJP} shows the $x(y)$ and $x_{\rm rip}(y)$ data for the left edges of two laboratory grown icicles, one made from distilled water and the other from distilled water with $8.0\times10^{-3}$ wt~\% salt added.  

It is apparent from Fig.~\ref{fig_ripalg_NJP} that the ripples can have considerable variation in their peak heights. To characterize the amplitude and wavelength of the ripples, we calculated the Fourier power spectrum of the spatial series $x_{\rm rip}(y)$ for the top 10~cm of the left and right edges, which were spliced together peak-to-peak.  In the Fourier analysis, we included only ripples in the uppermost 10~cm of the icicle, because ripples farther down had had less time to grow and may grow under more impure conditions due to the exclusion of impurities by ice formation higher up.   The rightmost plots in Fig.~\ref{fig_ripalg_NJP} show the power spectra of the de-trended data.  We defined the ripple amplitude $A$ as the value of the maximum point in the Fourier spectrum, with a small background subtracted.  The background was determined by fitting a Lorentzian function plus a constant positive offset to the spectral data through the maximum point.  We defined the ripple wavelength $\lambda$ to be that corresponding to the wavenumber of the maximum point; its error was given by the half-width at half maximum of the Lorentzian fit.  This analysis extracts the amplitude and wavelength of the most dominant Fourier mode of the ripple pattern and ignores other features of its power spectrum.  Isolated bumps on the icicle surface with a length scale comparable to that of ripples may contribute to the power of the peak.  Since distilled water icicles are not perfectly smooth, they still have a broad but weak Fourier spectrum after de-trending. 

 The resolution of the digital images was typically 0.018~cm per pixel; thus, to detect the edge of one side of a 10~cm segment of the icicle required about 550 edge detection measurements.  Since we determined the ripple amplitude and wavelength using both the left and right edges, the analysis of a single image involved about 1100 edge detections. These data were further averaged over four rotational views of the icicle. The amplitude resolution limit was  approximately 0.01~cm, or about half a pixel.  The entire algorithm for ripple extraction and analysis was tested on simulated icicle images with the same resolution as that of the actual data.  The algorithm accurately yielded the correct known ripple amplitudes and wavelengths of the simulated images.  


\begin{figure}
\begin{center}
\includegraphics[width=15.5cm]{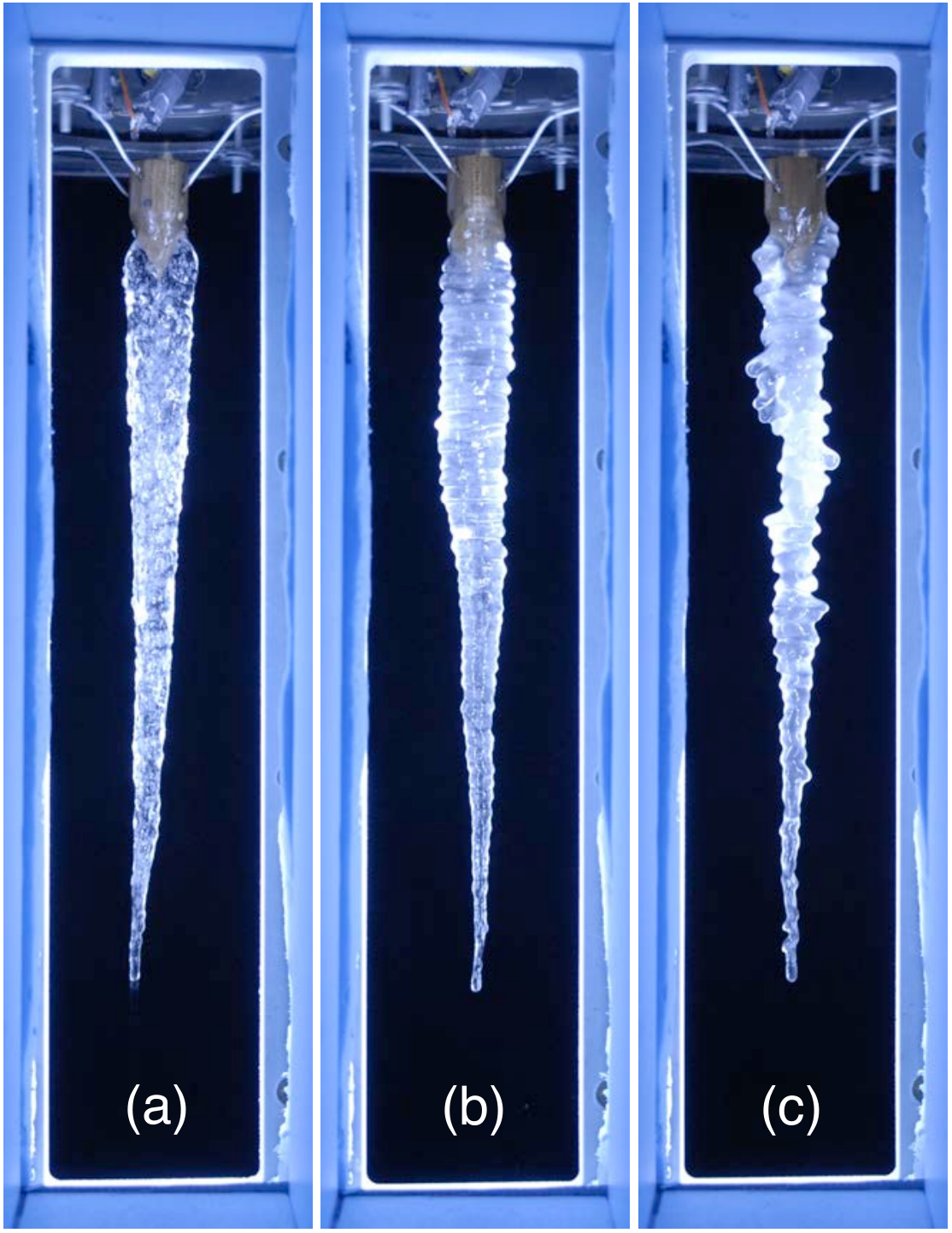}
\caption{\label{fig_ripplesvssalt_pics}
Images of three icicles grown under identical conditions: an ambient wall temperature of -12.3~$^\circ$C, a water supply rate of 2.0~g/min, a nozzle temperature of 3.0~$^\circ$C, and a volumetric air flux of 0.95~m$^3$/min per fan.  (a) was made with distilled water; (b) was made with distilled water plus $(8.0\pm0.2)\times10^{-3}$~wt~\% NaCl; (c) was made with distilled water plus $(1.28\pm0.02)\times10^{-1}$~wt~\% NaCl. 
}
%
%
%
%
\end{center}
\end{figure}

\begin{figure}
\begin{center}
\includegraphics[width=15.5cm]{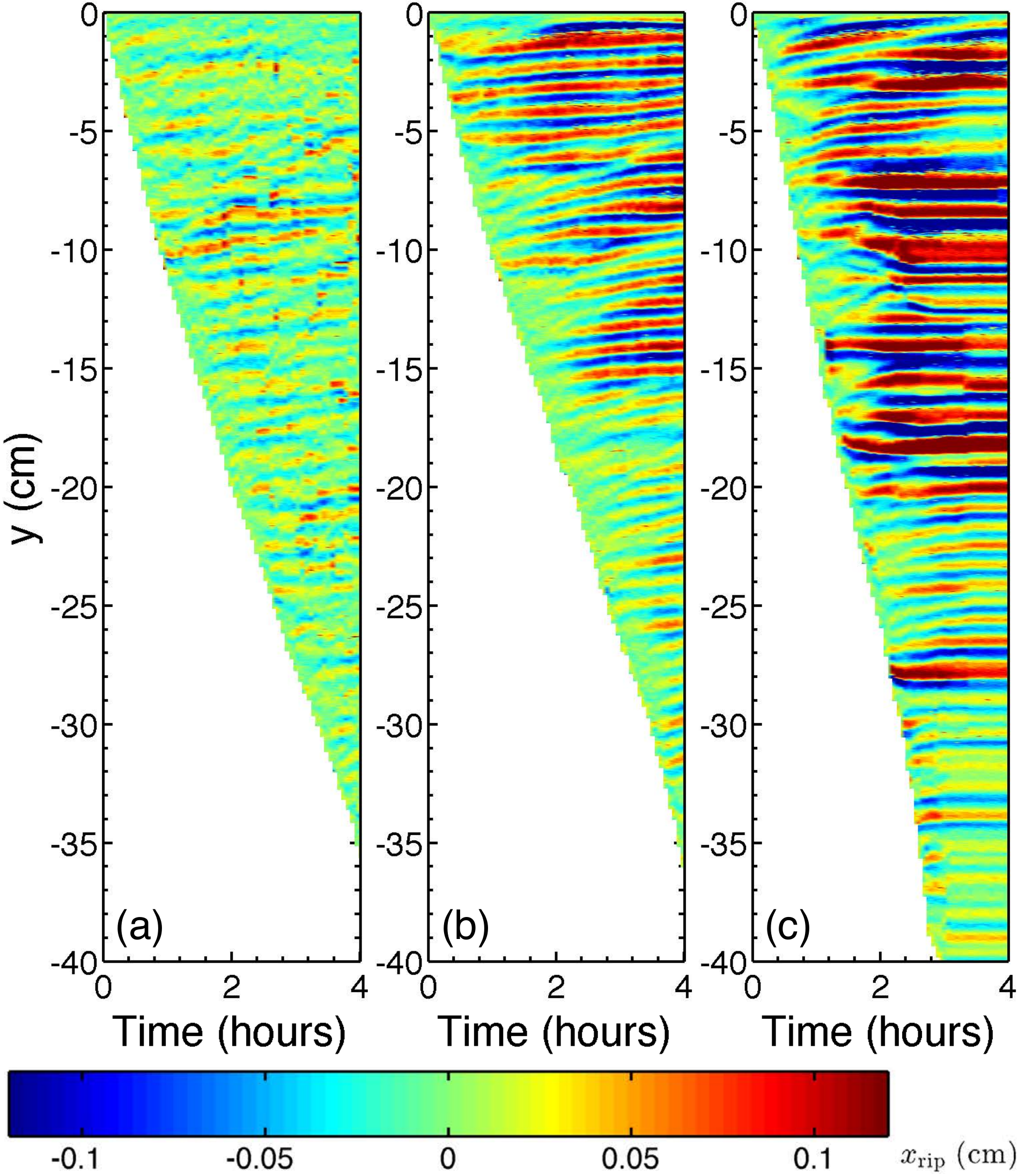}
\caption{\label{fig_ripplesvssalt_spacetime}
Space-time plots showing the evolution of the ripple topography, viewed from one fixed rotational position, for the right edges of the three icicles shown in Fig.~\ref{fig_ripplesvssalt_pics}.  $x_{\rm rip}$, indicated by the color, is plotted as a function of $y$ vs. time.  The border of the white region follows the downward growth of the icicle tip.  The slow upward motion of the ripples is apparent in the small positive slope of the crests, especially in part (b).}
\end{center}
\end{figure}

\section{Results}


In all, we analyzed the ripples on 67 laboratory icicles, grown under a broad range of conditions. We first focus on comparing icicles made from feed water with different compositions, grown under otherwise identical conditions: an ambient wall temperature of $-12.3\pm0.2$~$^\circ$C, a water supply rate of $2.0$~g/min, a nozzle temperature of $3.0\pm0.3$~$^\circ$C, and a volumetric air flux of $0.95\pm0.03$~m$^3$/min per fan.  Fig.~\ref{fig_ripplesvssalt_pics} displays three such icicles made from distilled water plus varying amounts of NaCl.  Even from direct inspection, there are notable differences between them.  The distilled water icicle is close to the self-similar shape~\cite{Chen_PRE,Short_Phys_Fluids} and has a relatively smooth surface.  When a sufficient amount of salt is added to the water supply, measurable ripples emerge on the surface of the icicle, and its overall form deviates from self-similarity and possibly also axisymmetry.  As the salt concentration is increased, as in Fig.~\ref{fig_ripplesvssalt_pics}(c), the ripples become more significantly non-sinusoidal and the overall shape of the icicle becomes more distorted.

The time evolution of the ripple pattern can be visualized by plotting the icicle topography in space-time. Fig.~\ref{fig_ripplesvssalt_spacetime} shows space-time plots of the right-hand edges of the same three icicles shown in the previous figure.  The time evolution of $x_{\rm{rip}}(y)$ is plotted for a fixed rotational position.  Some small features that survive the de-trending filter are found even on the distilled water icicle.  Saltier icicles show clear ripples that appear in a patchy way and sometimes quickly grow to a saturated amplitude which can be as large as a few millimetres.  Slow upward ripple motion, which was previously observed using tap water~\cite{ueno_ripple_expt,Chen_PRE}, can be seen in parts of the ripple  patches.  The overall speed of this motion is consistent with previous observations~\cite{ueno_ripple_expt,Chen_PRE}.  The upward motion is sometimes locally interrupted by the appearance of new ripples, ripple mergers, and other dynamics.

~~

\begin{figure}
\vspace{1.6cm}
\begin{center}
\includegraphics[width=15.5cm]{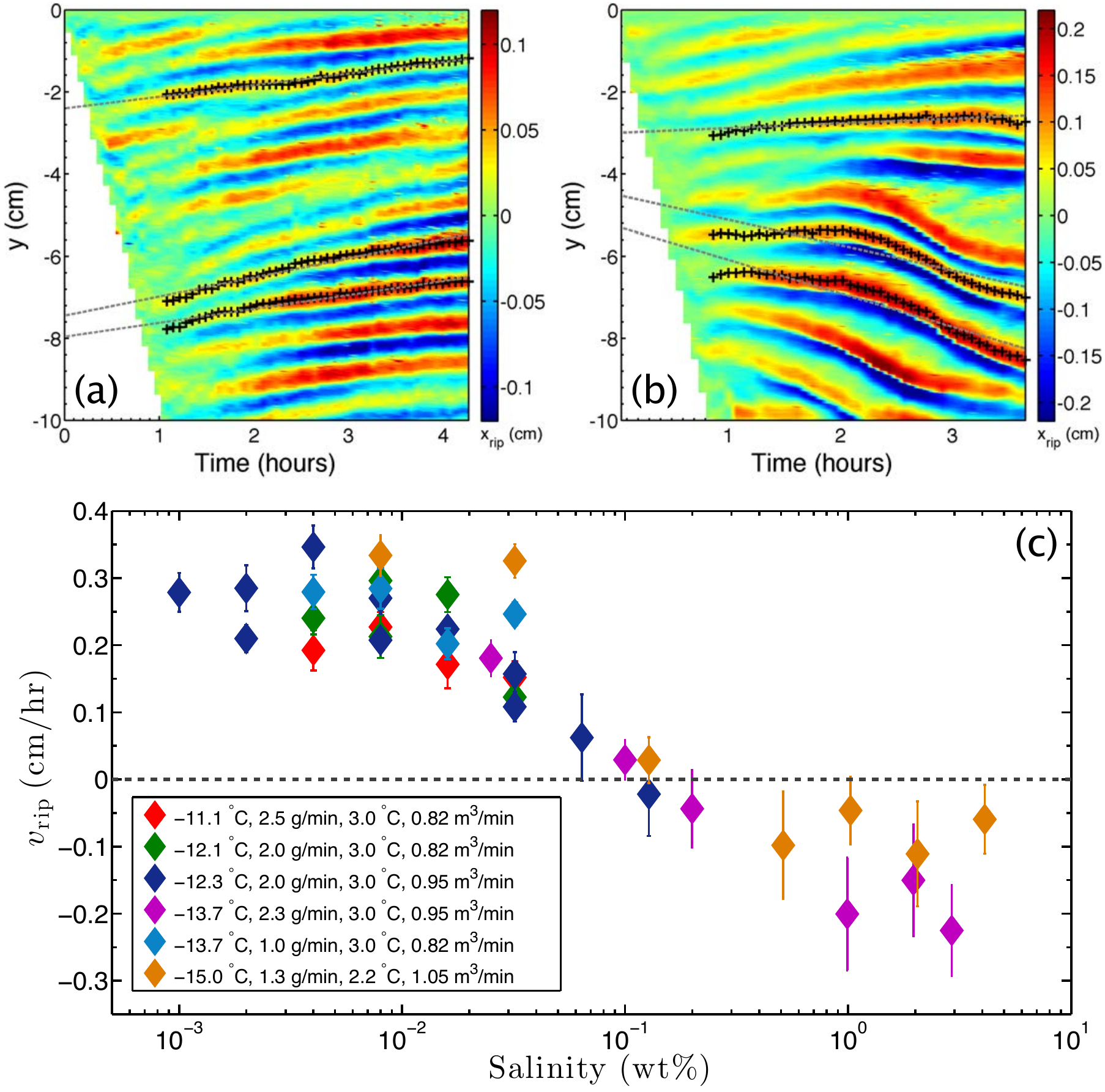}
\caption{\label{fig_rip_vs_salinity_abc}
Two space-time plots showing the evolution of the ripple topography of one edge, viewed from one fixed rotational position, for the top 10~cm of saline icicles with (a) $4.0\times10^{-3}$~wt~\% and (b) $6.4\times10^{-2}$~wt~\%  salt.  Note the difference in the scales on the color bars. The crosses follow the peaks of three randomly selected ripples in time, starting when the icicle reached 10~cm in length.  The dashed lines are best linear fits through the peak positions; their slope gives the average ripple traveling velocity $v_{\rm{rip}}$, with positive upward.  (c) Ripple traveling velocities $v_{\rm{rip}}$ \emph{vs.} feed water salinity, with ambient wall temperature, water supply rate, nozzle temperature, and fan speed fixed. }
\end{center}
\end{figure}

To characterize the velocity of the ripple motion on an icicle for a fixed rotational position, three ripple peaks were randomly identified, starting 
when the icicle reached 10~cm in length.  The ripple peak positions $y_{\rm{rip}}(t)$ were followed through time, and their phase speeds $v_{\rm{rip}}$ were extracted from the slope of best-fit straight lines.  Two applications of this procedure are illustrated in Fig.~\ref{fig_rip_vs_salinity_abc}(a) and (b). 
 In Fig.~\ref{fig_rip_vs_salinity_abc}(c), we show $v_{\rm{rip}}$ as a function of the feed water salinity for several sets of conditions.  Here, 24 speed measurements have been averaged over one icicle rotation, and the error bars show the standard errors of the mean.  Below about 0.05~wt~\%, the ripples  climbed the icicle uniformly at speeds of millimetres per hour, as shown in Figs.~\ref{fig_ripplesvssalt_spacetime}(b) and~\ref{fig_rip_vs_salinity_abc}(a).   Above 0.05~wt~\%, uniform ripple motion was less frequently observed, as the ripple dynamics became more complex, as in Figs.~\ref{fig_ripplesvssalt_spacetime}(c) and and~\ref{fig_rip_vs_salinity_abc}(b).  Some ripples moved nonlinearly, and different ripples on the same icicle may travel in opposite directions, as in Fig.~\ref{fig_rip_vs_salinity_abc}(b).  This may result in an average $v_{\rm{rip}}$ that is negative and contributes to a larger uncertainty in $v_{\rm{rip}}$.  The ripple speeds were not strongly dependent on the ambient temperature, input mass flux, fan speed, or the nozzle temperature.  The linear stability theory of Ueno {\it et al.}~\cite{ueno1,ueno2,ueno4,ueno5,ueno_ripple_expt,ueno_farz_2011}  predicts an upward ripple motion of about the right magnitude, while Ogawa {\it et al.}~\cite{ogawa} predict downward motion.  Neither theory accounts for salinity, however.

\begin{figure}
\begin{center}
\includegraphics[width=15.5cm]{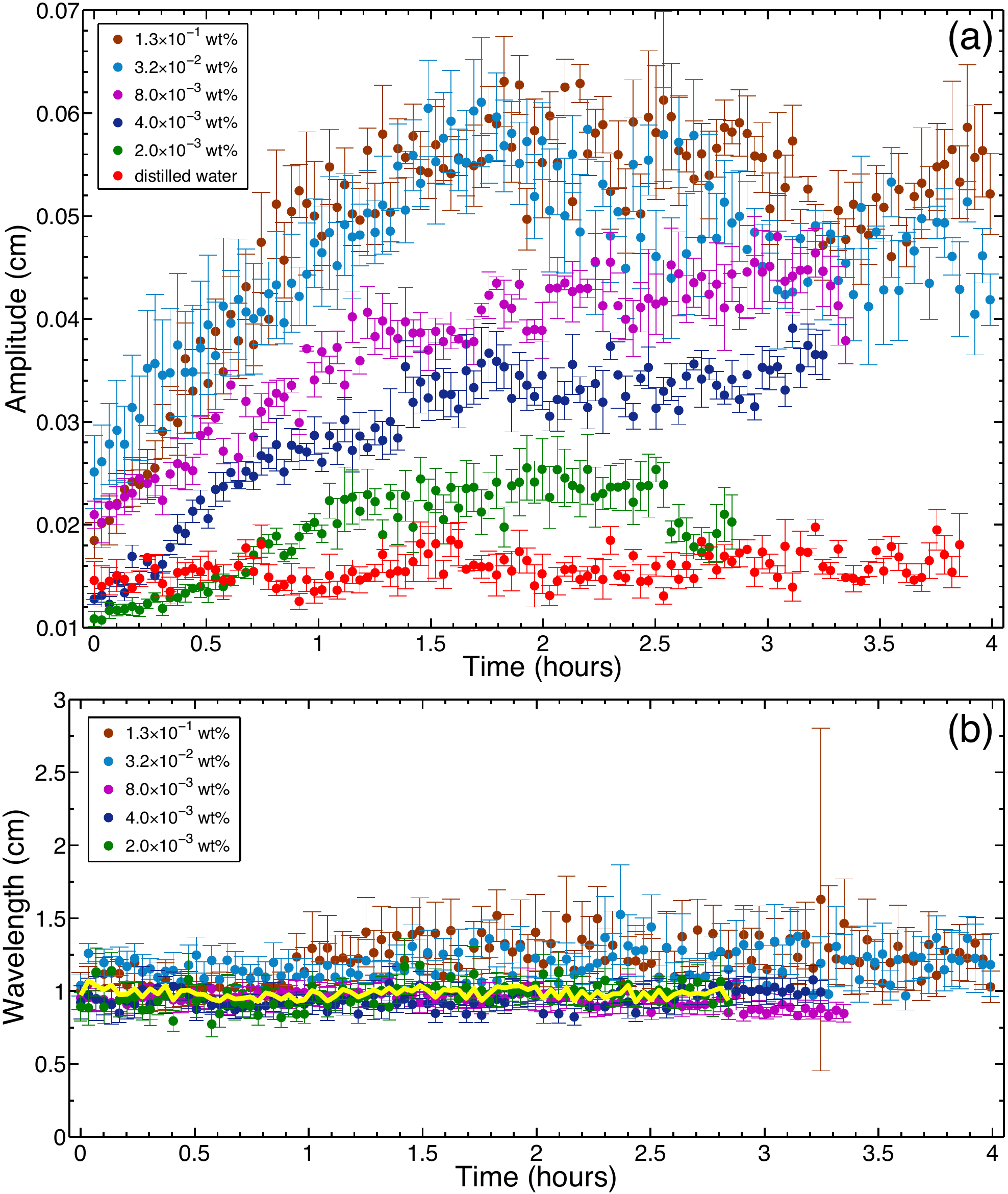}
\caption{\label{fig_ripplesvstime_salt}
Time-series of the (a) amplitude and (b) wavelength of ripples on icicles made from feed water with varying salinity, which is indicated in the legend.  In all experiments, the ambient wall temperature was -12.3~$^\circ$C, the water supply rate was 2.0~g/min, the nozzle temperature was 3.0~$^\circ$C, and the surrounding air flux due to each fan was 0.95~m$^3$/min.  Time zero corresponds to the time when the icicle reached 10~cm in length.  The error bars in (a) indicate the standard error from averaging over each half-rotation.  The half widths at half-maximum from the Lorentzian fits were used to estimate the errors in (b).  The yellow line in (b) shows the error-weighted mean of the wavelength as a function of time.}
\end{center}
\end{figure}

We now turn to the time evolution of the amplitude of the ripples.  Here we find that the salinity is the controlling parameter for the existence or non-existence of growing ripples. Fig.~\ref{fig_ripplesvstime_salt} shows the time evolution of the ripple amplitude and wavelength for icicles made from feed water with varying salinity.  These were obtained for each image using the Fourier methods described above.  Here, time zero corresponds to the time when the icicle reached 10~cm in length, and the data has been averaged over each half-rotation, which consists of four unique views; thus, each data point is based on about 4400 edge detection measurements.  The amplitudes in Fig.~\ref{fig_ripplesvstime_salt}(a) are smaller than typical amplitudes within the patches of ripples in Fig.~\ref{fig_ripplesvssalt_spacetime}, because the Fourier analysis included regions of the icicle with small or no ripples.  The error bars, estimated from the standard error of the averaging sample, are larger for saltier icicles; this reflects the progressive loss of axisymmetry as the feed water salinity was increased.

As seen in Fig.~\ref{fig_ripplesvstime_salt}(a), throughout the growth of the distilled water icicle with no added salt, the amplitude of the residual topographic features that survived the de-trending filter remained a few times the amplitude resolution limit.  In contrast, saline water icicles exhibited clear ripples, whose amplitudes  increased to  significant values before saturating. Some amplitudes peaked or plateaued but later decayed, particularly in the more saline cases.   The ripples on more saline icicles grew faster and reached higher amplitudes.

\newpage


Ripple growth was not observed on distilled water icicles. Measurable ripple growth became apparent at the remarkably low feed water salinity of $2.0\times10^{-3}$~wt~\%, {\it i.e.} only 20~mg of salt per litre of water.  This is, in fact, a considerably lower level of ionic impurities than in typical tap water, as shown in Table~\ref{waters}. These very small ionic impurity levels can also be reached or exceeded in natural icicle source water, also shown in Table~\ref{waters}.  This shows why at least some natural icicles can become ripply under normal outdoor conditions, even without any obvious or unusual  source of the impurities.  

 
Even while ripples were growing, they maintained a very constant wavelength. The error-weighted mean of the ripple wavelength as a function of time for various saline water icicles is plotted  in Fig.~\ref{fig_ripplesvstime_salt}(b).   
Its time average is $0.985\pm0.004$~cm.  The ripple wavelength is essentially independent of time and hence of the ripple amplitude.

\begin{figure}
\begin{center}
\includegraphics[width=15.5cm]{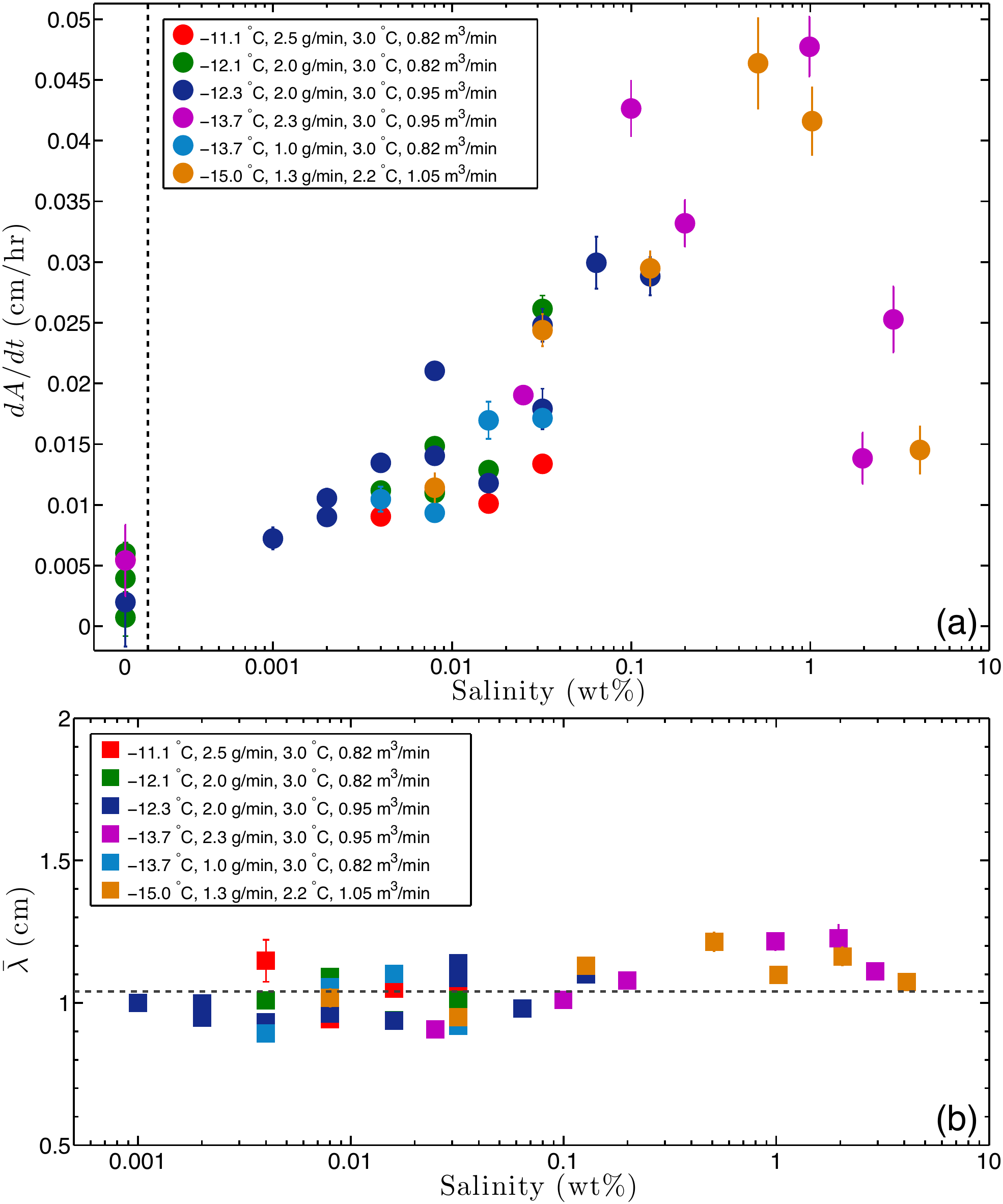}
\caption{\label{fig_rippleresults_vssalinity}
(a) Amplitude growth speeds $dA/dt$ and (b) time-averaged wavelengths  $\bar{\lambda}$ of icicle ripples \emph{vs.} feed water salinity, under various conditions of ambient wall temperature, water supply rate, nozzle temperature, and air flux per fan.  Each color corresponds to one set of experiments with the same values of the extrinsic conditions.  The vertical dashed line in (a) is a scale break, to the right of which salinity is plotted on a log scale.  The horizontal dashed line in (b) indicates the global mean of $\bar{\lambda}$.}
\end{center}
\end{figure}

To compare ripples grown under diverse conditions, we calculated the amplitude growth speed $dA/dt$ and the time-averaged wavelength $\bar{\lambda}$ of the time series as given by Fourier analysis.  We considered ripples grown under various ambient temperatures, input mass flux, nozzle temperatures, and air flow. In each case, to find $dA/dt$, we smoothed the raw $A(t)$ data, and performed a linear fit within the linear regime of ripple growth, defined to be from $t=0$ to the time at which the slope decreases by more than 50~\%.  The time-averaged wavelength $\bar{\lambda}$ was found for the same linear growth regime.  Again, salinity proved to be the only significant controlling parameter. Fig.~\ref{fig_rippleresults_vssalinity}(a) shows the ripple growth speed $dA/dt$ plotted as a function of the feed water salinity for several sets of conditions.  As the salt concentration of the water supply was increased from zero, the ripples on the resultant icicle grew more rapidly, but the dependence was extremely weak.  The increase is initially approximately logarithmic; for a thousandfold increase in salt concentration, $dA/dt$ only increases by about a factor of three.  For salinity above 1~wt~\%, the trend in $dA/dt$ exhibits more scatter and an overall negative correlation with salinity.  On the other hand, as shown in Fig.~\ref{fig_rippleresults_vssalinity}(b), the time averaged wavelength $\bar{\lambda}$ remains robustly independent of the feed water salinity, although the saltiest icicles all had wavelengths that slightly exceeded 1~cm.  The global mean of $\bar{\lambda}$ was $1.04\pm0.01$~cm.


\begin{figure}
\begin{center}
\includegraphics[width=15cm]{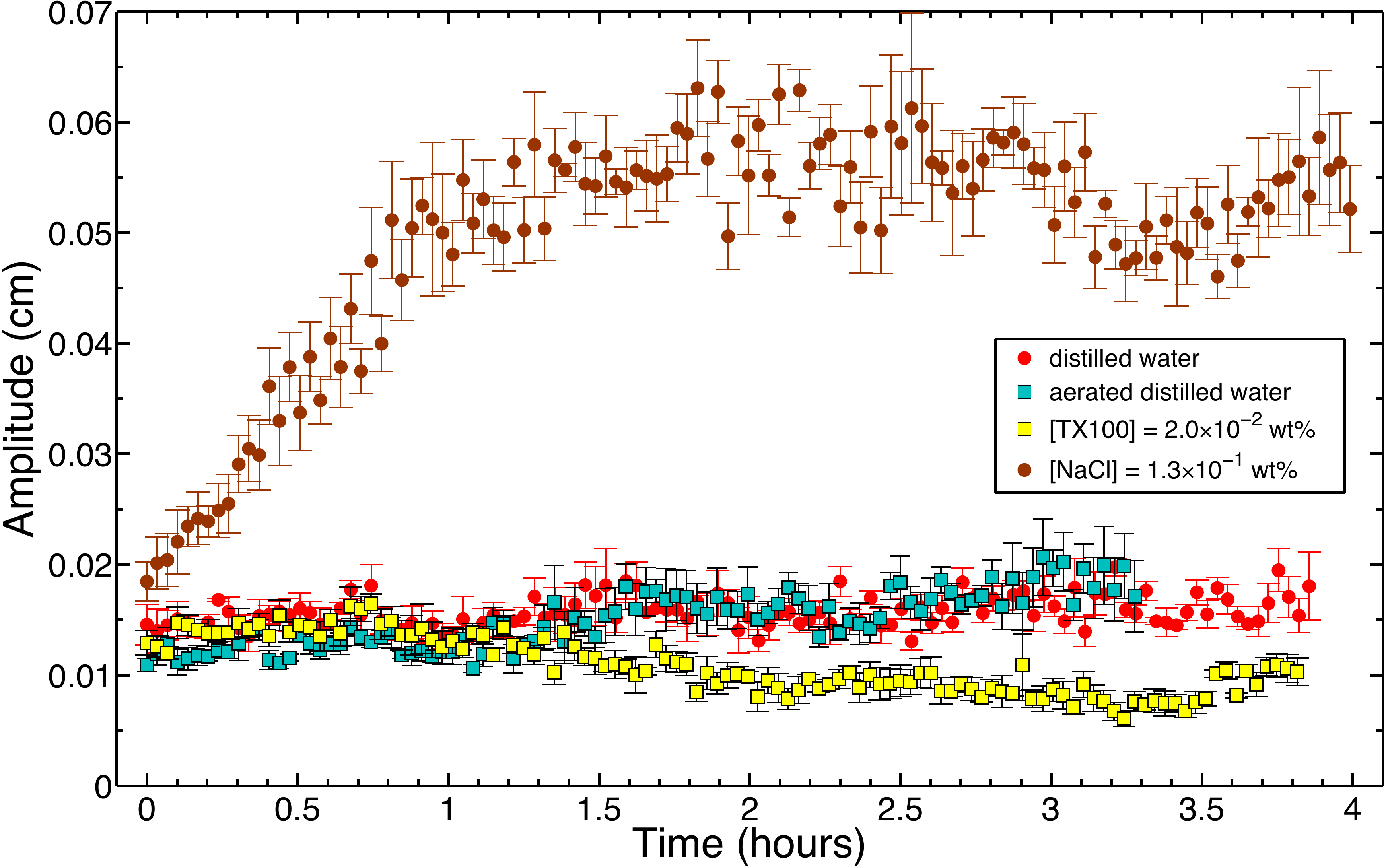}
\caption{\label{fig_ripplesvstime_others}
Time-series of the ripple amplitude on icicles made from various solutions of salt, air, and surfactant, dissolved in distilled water.  For all experiments, the ambient wall temperature was -12.3~$^\circ$C, the water supply rate was 2.0~g/min, the nozzle temperature was 3.0~$^\circ$C, and the surrounding air flux was 0.95~m$^3$/min per fan.  The error bars indicate the standard error from averaging over each half-rotation.  Neither dissolved air nor dissolved non-ionic surfactant alone produced ripples.}
\end{center}
\end{figure}

Given the very sensitive dependence of the ripple growth on dissolved impurities, it is natural to examine the effect of dissolved gases.  All of the water used in our experiments had been exposed to air and presumably contained some concentration of dissolved oxygen and other gases.  This is of course also true of water that forms natural icicles.  To study the effect of dissolved gases, we grew icicles from distilled water that had air bubbled through it for 12 hours, which may therefore be assumed to be saturated, or possibly supersaturated, with gases.   Fig.~\ref{fig_ripplesvstime_others}(a) includes a typical time-series of the ripple amplitude for an icicle made from deliberately aerated distilled water.  The resulting icicles were indistinguishable from those grown from standard distilled water in that no growing ripples were formed.  We conclude from this that dissolved gases alone are insufficient to trigger the rippling instability.

~~

~~


As discussed in the next section, some theories of the rippling instability~\cite{ueno1,ueno2,ueno4,ueno5,ueno_ripple_expt,ueno_farz_2011} predict that the water-air surface tension $\gamma$ is an important parameter.  In order to test this idea, we employed the non-ionic surfactant Triton X-100 to vary the surface tension.  As shown in Table~\ref{waters}, this surfactant has a negligible effect on the conductivity of the solution. We grew icicles from solutions of distilled water and sufficient  Triton X-100 to reduce the surface tension by more than 45~\%.  As shown in Fig.~\ref{fig_ripplesvstime_others}, these icicles again had an amplitude equal to, or even below, that of standard distilled water icicles and exhibited no measurable ripple growth.  Thus, reduced surface tension alone does not cause ripples to grow.



\section{Discussion}



Models of icicles~\cite{ogawa,ueno1,ueno2,ueno4,ueno5,ueno_ripple_expt,Short_Phys_Fluids,makkonen1} assume that their surface is covered by a thin film of flowing water which has a parabolic shear flow profile with a surface speed $U$.  Film thicknesses $h$ are typically 100\,$\mu$m or less, and their flow is laminar.  Any topography present on the ice surface is reflected in the shape of the water-air interface in a manner controlled by the surface tension $\gamma$.  Because the thermal conductivity of air is much less than that of water, the temperature difference $\Delta T $ across the thin film is much smaller than the total temperature difference driving the growth~\cite{ueno4,Neufeld_JFM}.  In our experiments, the latter was $\sim$ 10$~^\circ$C, while  $\Delta T < 0.1~^\circ{\rm C}$.  The latent heat released at the ice-water interface as ice is formed is diffused and advected through the flowing water film and carried away by the surrounding stirred air.  The relevant dimensionless parameter characterizing the relative importance of advection and diffusion in the water film is the  P{\'e}clet number ${\rm Pe}=  U  h/ \kappa \sim 5$, where $\kappa$ is the thermal diffusivity of water.
   It has been shown that, in addition to diffusion and advection, both thermal radiation and water evaporation with vapor advection by the air make significant contributions to the heat transport outside the icicle~\cite{Neufeld_JFM}.

Existing theories of icicle ripples~\cite{ogawa,ueno1,ueno2,ueno4,ueno5,ueno_ripple_expt} have not considered the effect of impurities, and included only thermal and surface tension effects.   The linear stability analysis by Ueno~\cite{ueno4} found an instability to growing ripples that travel upward slowly, against the direction of flow.  The most fundamental difficulty with this theory is that it predicts growing ripples on pure water icicles, which we do not observe.  No current theory predicts the salt concentration dependence of the ripple growth speed shown in Fig.~\ref{fig_rippleresults_vssalinity}(a).

Furthermore, the theory of Ueno~\cite{ueno4} does not correctly account for the effect of surface tension, with or without added salt or surfactant.  The predicted wavenumber $k_c$ and exponential amplification rate $\sigma_r$ of the most unstable mode are given by~\cite{ueno4}
\begin{equation}
k_c  = \bigg[\frac{3~\rho~g}{{\rm Pe}~\gamma~h}\bigg]^{1/3}, ~~~~~ \sigma_r(k_c) = \frac{3 }{4}
~V ~k_c~\sim \gamma^{-1/3}, \label{ueno_predicts}
\end{equation}
where $\rho$ is the density, $g$ is the acceleration due to gravity, and $V$ is the mean radial growth speed.  Thus,  an increase in surface tension is predicted to result in a weak decrease in the ripple amplification rate.  The effect of added salt on the surface tension is much too small and of the wrong sign to account for the growth of the ripples.  Over the range of salt concentrations in Fig.~\ref{fig_rippleresults_vssalinity}(a), the air-water surface tension increases by 0.4~\%~\cite{MIT_seawater}, while the measured amplification rate, which we obtained by a fitting an exponential  to the initial rise of $A(t)$, increases by a factor of 5.  The addition of surfactant, which greatly lowers the water-air surface tension, produced no measurable ripples and hence no increase in their growth rate.  These observations contradict the $\gamma^{-1/3}$ dependence of Eqn.~\ref{ueno_predicts}.  Thus, our results are broadly inconsistent with existing theories~\cite{ogawa,ueno1,ueno2,ueno4,ueno5,ueno_ripple_expt,ueno_farz_2011}.  A theory of the rippling instability that is consistent with all the experimental facts is currently lacking.

The intriguing similarity between icicle ripples and stalactite crenulations~\cite{stalactite_ripples} may offer some new insights.   A fresh theoretical approach to icicle ripples must include, at the very least, the advection and diffusion of both heat and salt in the water film.  The analogy between icicles and stalactites is not perfect, however~\cite{Short_Phys_Fluids,Short_PRL_stalactites}.  Stalactites are formed by the deposition of CaCO$_3$ under similar thin-film flow conditions, and also involve several advecting and diffusing species, with [CO$_2$] playing a similar role to the temperature in icicles.  But the dynamics of stalactite formation is complicated by liquid-phase chemical reactions that have no analog in saline icicles~\cite{stalactite_ripples}.

~~

Moreover, the ice-water interface of an icicle is unstable at small scales to the formation of dendrites when freezing proceeds into the supercooled water film~\cite{MS1,MS2}.  Under typical experimental conditions, Mullins-Sekerka theory~\cite{MS1,MS2} predicts a dendritic spacing of 0.006~cm, much smaller than the typical wavelength of icicle ripples.  During such ``wet" ice growth~\cite{wet_ice_growth}, a significant fraction of liquid water, whose freezing point is depressed by the presence of salt, can be trapped between the advancing dendrites~\cite{wet_ice_growth,makkonen1,SF2,ueno_farz_2011}.  The dynamics of such a ``mushy layer"~\cite{mushy1,mushy2} under a \emph{thin} flowing water film is not well-understood, and may explain the impurity dependence of the ripple formation we observed.  Future experiments should investigate the coupled effect of ionic impurities and surface tension, by growing icicles using solutions containing various types of both salt and surfactant.  In addition to morphological analysis, measuring the evolving composition of the solid phase of the icicle and of the drip-off water may reveal further insights into the mechanism of the rippling instability.

\section{Conclusion} 


We have shown experimentally that small amounts of ionic impurities are required for the formation of ripples on icicles and that growing ripples are not present on icicles made from pure water, even with dissolved gases.  Growing ripples are not produced by simply reducing the surface tension of pure water with a surfactant.  Ripple growth rates show a very weak, roughly logarithmic dependence on ionic impurity concentration. With impurities present, the ripple wavelength remains constant under all other variations of the growth conditions.  The speed and even the direction of ripple motion depend on the impurity concentration. Existing theories of ripple formation do not account for the effect of impurities and cannot be easily generalized to include them. Thus, while our experiments have provided much new information on the origin of icicle ripples, this elegant example of natural pattern formation remains theoretically unexplained.
 

\section{Acknowledgements}

%
%

We thank J. S. Wettlaufer, J. A. Neufeld, M. G. Worster, R. E. Goldstein, and K. Ueno for insightful discussions, and C. Ward for his assistance with the surface tension measurements.  This work was supported by the Natural Sciences and Engineering Research Council of Canada.

\end{document}